\documentclass[aps,prl,twocolumn]{revtex4}

\newcommand{\I}{\mathrm{i}}
\newcommand{\expect}[1]{\bigl\langle{#1}\bigr\rangle}
\newcommand{\tr}[1]{\mathrm{tr}\bigl\{#1\bigr\}}
\newcommand{\ket}[1]{| #1 \rangle}
\newcommand{\bra}[1]{\langle #1 |}

\newcommand{\Exp}[1]{\mathrm{e}^{\mbox{\footnotesize$#1$}}}

\begin{document}
\title{Comment on ``Minimum Uncertainty and Entanglement''}

\author{Berthold-Georg Englert}
\affiliation{Centre for Quantum Technologies, %
National University of Singapore, Singapore 117543, Singapore}
\affiliation{Department of Physics, %
National University of Singapore, Singapore 117542, Singapore}

\date[Posted on ]{4 August 2011; updated on 3 September 2011}

\begin{abstract}
Dass, Qureshi, and Sheel conjecture that the lower bound in the
Heisenberg--Robertson uncertainty relation cannot be reached in mixed
states. The conjecture is wrong.
\end{abstract}

\maketitle

The Heisenberg--Robertson uncertainty relation,
\begin{equation}
  \label{eq:1}
  \delta A\;\delta B\geq\frac{1}{2}\Bigl|\expect{\I\bigl[A,B\bigr]}\Bigr|\,,
\end{equation}
sets a lower bound on the product of the spreads
\begin{eqnarray}
  \label{eq:2a}
  \delta A&=&\sqrt{\expect{A^2}-\expect{A}^2}\,,\nonumber\\
  \delta B&=&\sqrt{\expect{B^2}-\expect{B}^2}
\end{eqnarray}
of two hermitian observables $A$ and $B$, where the expectation values refer
to a state specified by a statistical operator $\rho$, as exemplified by
\begin{equation}
  \label{eq:2b}
  \expect{A}=\tr{A\rho}\,.
\end{equation}
All of this is familiar textbook fare, of course; see, for example,
Sec.~4.7 in \cite{SimpSys}.

First Sheel and Qureshi \cite{SQ-arXiv} 
and then Dass, Qureshi, and Sheel \cite{DQS-arXiv} conjectured that, for 
${\delta A\;\delta B>0}$, the lower bound in (\ref{eq:1}) can only be reached
if the state is pure, that is $\rho^2=\rho$.
The following simple counter example demonstrates that the conjecture is wrong.

Consider the hermitian observables
\begin{equation}
  \label{eq:3a}
  A=\bigl(\ket{1},\ket{2},\ket{3}\bigr)
    {\left(\begin{array}{ccc}0&1&0\\1&2&0\\0&0&0
           \end{array}\right)}\!
    {\left(\begin{array}{@{}c@{}}\bra{1}\\\bra{2}\\\bra{3}
           \end{array}\right)}   
\end{equation}
and
\begin{equation}
  \label{eq:3b}
  B=\bigl(\ket{1},\ket{2},\ket{3}\bigr) 
    {\left(\begin{array}{ccc}0&-\I&0\\\I&0&0\\0&0&0
           \end{array}\right)}\!
    {\left(\begin{array}{@{}c@{}}\bra{1}\\\bra{2}\\\bra{3}
           \end{array}\right)}   
\end{equation}
together with the mixed-state statistical operator \cite{mixed}
\begin{equation}
  \label{eq:3c}
  \rho=\bigl(\ket{1},\ket{2},\ket{3}\bigr)
    \frac{1}{2}{\left(\begin{array}{ccc}1&0&0\\0&0&0\\0&0&1
          \end{array}\right)}\!
    {\left(\begin{array}{@{}c@{}}\bra{1}\\\bra{2}\\\bra{3}
           \end{array}\right)},   
\end{equation}
where $\ket{1}$, $\ket{2}$, $\ket{3}$ are three kets from an orthonormal basis,
and $\bra{1}$, $\bra{2}$, $\bra{3}$ are the corresponding bras.
For these, we have
\begin{equation}
  \label{eq:4}
  \bigl(\delta A\bigr)^2=\bigl(\delta B\bigr)^2=
\frac{1}{2}\Bigl|\expect{\I\bigl[A,B\bigr]}\Bigr|=\frac{1}{2}\,,
\end{equation}
and the lower bound in (\ref{eq:1}) is reached, indeed.

It is easy to construct more counter examples. As the standard derivation of
(\ref{eq:1}) shows, the main ingredient is an operator $A+\I B$ that is not
normal and has degenerate eigenvalues.

The alleged proof of the conjecture in \cite{DQS-arXiv} relies crucially on the
wrong assertion \cite{DQS-details} 
that eigenkets of $A+\I B$ are necessarily simultaneous eigenkets of $A$ and
$B$. In the example above, ket $\ket{1}$ is an eigenket of
$A+\I B$, but it is not an eigenket of $A$ or of $B$.

The single counter example of (\ref{eq:3a})--(\ref{eq:4}) should suffice.
But, just in case, here are two more counter examples, 
one for angular-momentum states, 
the other for gaussian states, two of the situations considered
in \cite{SQ-arXiv,DQS-arXiv}.

For the angular momentum vector operator $\mathbf{J}$ with cartesian
components $J_x$, $J_y$, $J_z$, we denote the joint eigenkets of
$\mathbf{J}^2$ and $J_z$ by $\ket{j,m}$ as usual. 
The equal sign holds in (\ref{eq:1}) for the pair of observables 
$A=J_x$, $B=J_y$ and the mixed-state statistical operator
\begin{equation}
  \label{eq:5}
\rho=\frac{1}{2}\bigl(\ket{0,0}\bra{0,0}+\ket{1,1}\bra{1,1}\bigr)\,.  
\end{equation}

For a Heisenberg pair $X$, $P$ with $[X,P]=i\hbar$, we denote by $\ket{a}$ the
ket with position wave function 
\begin{equation}
  \label{eq:6a}
\psi_a(x)=(\kappa/\pi)^{1/4}\,\Exp{-\frac{1}{2}\kappa(x-a)^2},   
\end{equation}
where $a$ is real and $\kappa$ is a fixed positive constant.
The equal sign holds in (\ref{eq:1}) 
for the pair of observables 
\begin{equation}
  \label{eq:6b}
A=XP+PX\,,\quad B=(\hbar\kappa X)^2-P^2
\end{equation}
and the mixed-state statistical operator
\begin{equation}
  \label{eq:6c}
\rho=\frac{1}{2}\bigl(\ket{a}\bra{a}+\ket{-a}\bra{-a}\bigr)  
\end{equation}
with $a\neq0$.


\begin{thebibliography}{9}
\bibitem{SimpSys} 
B.-G. Englert, \textit{Lectures on quantum mechanics---Basic matters\/} (World
Scientific, Singapore 2006).

\bibitem{SQ-arXiv}
A. Sheel and T. Qureshi, e-print arXiv:1107.5929v1.

\bibitem{DQS-arXiv}
N.D.H. Dass, T. Qureshi, and A. Sheel, e-print arXiv: 1107.5929v2.

\bibitem{mixed}
The mixed state could refer to a subsystem of a larger system that is 
in a pure state, as in the situations examined in \cite{SQ-arXiv,DQS-arXiv}, 
but it is irrelevant whether there really is such a larger system. 

\bibitem{DQS-details}
See the paragraph after equation (29) and the sentence after equation (30) in
\cite{DQS-arXiv}.  

\end{thebibliography}
\end{document}